%% file: 0_main.tex
\documentclass[manuscript,screen]{acmart}

\renewcommand\footnotetextcopyrightpermission[1]{}

\usepackage[T1]{fontenc}

\usepackage{amsfonts}
\usepackage{algorithmic}
\usepackage{graphicx}
\usepackage{textcomp}
\usepackage{xcolor}
\usepackage{verbatim}
\usepackage{listings}
\usepackage{multirow}
\usepackage{subcaption}
\usepackage{enumitem}
\usepackage{framed}
\usepackage{colortbl}
\usepackage{soul}
\usepackage{tcolorbox}

\definecolor{lightgray}{gray}{0.95}
\definecolor{darkblue}{rgb}{0,0,0.5}
\definecolor{lightred}{rgb}{1, 0.8, 0.8}
\definecolor{lightyellow}{rgb}{1, 1, 0.8}
\definecolor{lightgreen}{rgb}{0.6, 1, 0.6}
\definecolor{mint}{rgb}{0.74, 0.99, 0.79}
\definecolor{forestgreen}{rgb}{0.13, 0.55, 0.13}

\AtBeginDocument{%
  }

\begin{document}

\title{The Prevalence and Impact of Licenses in Open Software Projects} 

\author{Mahmoud Jahanshahi}
\email{mjahan@utk.edu}
\affiliation{
  \institution{University of Tennessee}
  \city{Knoxville}
  \country{USA}
}

\author{Bogdan Vasilescu}
\email{vasilescu@cmu.edu}
\affiliation{
  \institution{Carnegie Mellon University}
  \city{Pittsburgh}
  \country{USA}
}

\author{Audris Mockus}
\email{audris@utk.edu}
\affiliation{
  \institution{University of Tennessee}
  \city{Knoxville}
  \country{USA}
}

\begin{abstract}
  The terms of how publicly available source code can be used are
  dictated by its license. The license (or its absence), in turn,
  affects what code the project may reuse and how its code can be
  (re)used and may also affect external participation and overall
  activity of the project. We aim to better understand the general
  state of license distribution overall and within language
  ecosystems and to investigate if license changes are associated
  with a noticeable variations of project output. To accomplish that
  we identify licenses and license types for over 100M software
  projects and find that most do not contain any license, that
  permissive licenses represent the bulk of most licenses, and that
  permissive licensing is representing an increasing proportion of
  all licenses over time. Restrictive licenses are more likely to be retained, however. There is a great variation among language
  ecosystems with C-language strongly favoring restrictive
  licenses. The analysis of license change impact comparing activity
  within one year of the adoption of the initial and final licenses
  shows that the change from restrictive to permissive license varies
  with the ecosystem. C-language ecosystems show reduced activity
  while Python shows increased activity when comparing restrictive to
  permissive license transition. Our results demonstrate dramatic
  changes in license type prevalence over time and find that the
  effects of license changes may have opposite effects depending on
  the language ecosystem.  
\end{abstract}

\ccsdesc[500]{Software and its engineering~Software creation and management}
\ccsdesc[500]{General and reference~Empirical studies}

\keywords{Software License, Open Source Software, Open Source License, World of Code}

\maketitle

\input{1_introduction}
\input{2_background}
\input{3_methodology}

\input{4_results}
\input{5_limitations}
\input{7_conclusions}

\section{Data Availability}
The replication package, including used datasets and R code are available at:
\url{https://zenodo.org/records/15031139}.

\clearpage
\bibliographystyle{ACM-Reference-Format}
\bibliography{ref}

\clearpage
\appendix
\input{8_appendix}

\end{document}

%% file: 1_introduction.tex
\section{Introduction}\label{intro}

Open Source Software (OSS) is central to modern
software development, fostering innovation, knowledge exchange, and
the widespread adoption of libraries and tools.  The
licensing of OSS projects dictates how software can be used,
modified, and distributed. This, in turn, should affect various aspects of projects' performance, such as new joiners, coding activity, the choice of upstream projects to depend upon, or the use by downstream projects.  

Our first aim is to better understand the actual state of license
selection in public software projects, including the choice of no license. 
Such unlicensed repositories play a significant role in the broader
ecosystem, e.g., they contribute code substantially to larger projects~\cite{jahanshahi2024beyond} and they represent a significant portion of the 
massive training data for LLMs~\cite{yu2023codeipprompt,xu2024first,jahanshahi2025cracks},
emphasizing their importance in understanding reuse patterns.

The absence of licenses introduces legal uncertainties that may hinder
downstream reuse. 
The problem may be propagated further, as the active and popular
projects are often upstream dependencies for numerous projects
downstream.
Hence if such inappropriate use is detected, it can affect
a very large part of the OSS ecosystem, not unlike what happened with
\textit{leftpad}~\cite{chowdhury2021untriviality}.
Prior work found relatively low frequency of repositories with no
license (e.g. 21.12\% in~\cite{wu2024large} and 10.51\% 
in~\cite{cui2023empirical}), that, as we demonstrate,
greatly underestimates the proportion of projects with no license
due to heavily filtering based on size, maturity, popularity, and
activity.

In OSS projects, decisions are often made based on perceived supply chain
concerns, such as choosing to participate in a project that will
become popular, hoping to have a large number of direct or
transitive downstream users (and followers), selecting upstream dependencies
that appear healthy, responsive, and
active~\cite{socialCoding,choice23,ma2020methodology}, and so on.
License choice should have clear and far reaching effects on the supply
chains in terms of attracting and retaining participants, having
downstream dependents or users, and, of course on ways the code may
be (re-)used.
Despite this, we are not aware of any prior studies
that model the impact of license choice nor any comprehensive theories that would
explain it.

Our second aim is, therefore, to establish and test a preliminary theory
that helps explain the impact of license choice in open software. 
The theory could help explain some of the reasons why a large fraction 
of projects have no license, why some license violations (such as the use of
unlicensed code) appear to be tolerated, and what to expect when a developer (or an organization) changes a license.
Such a theory can potentially lead to better tools that help developers
choose the most suitable license according to their objectives,
increase the proportion of projects with a license (hence spurring
innovation that is presently inhibiting actors who do pay attention
to licensing terms from using unlicensed code), and, more generally,
reduce risks arising from license incompatibilities in open software supply
chains.

To accomplish our goals we start from reviewing literature in areas
that may help us formulate a preliminary theory, such as OSS developer
motivation, company involvement in OSS, existing theories on how
technology and libraries are selected and spread, as well as
literature on OSS licensing. 
We pay particular attention to phenomena that could be measured at
scale in order to make our theory easily testable.
We then utilize a curated list of OSS projects with identified licenses 
and license types for nearly the entire OSS ecosystem in order 
to capture not only the most
popular projects but to include small and inactive projects that are
less likely to have a license. 
Based on this large curated list of
licenses and projects, we use our preliminary theory to
operationalize license choice where theory predicts a certain effect on 
key project factors.
We first test if certain types of licenses have longer
retention rates based on predictions of our theory.  We then employ
a multivariate multiple regression model
to test which of the theory-predicted contextual factors are impacted by
license choice.

Our analysis of the types and prevalence of licenses across more
than 131 million OSS projects, finds that 83\% of these projects
lack a formal license. Among the licensed projects, the MIT license
was the most commonly used, accounting for 65\% of all licensed
projects, followed by the Apache 2.0 and ISC licenses.  Although
permissive licenses dominate overall, copyleft licenses like the GPL
exhibited high retention rates, reflecting a strong commitment to
software freedom within certain communities. In comparison to prior
studies we find four-fold larger prevalence of unlicensed
code.

%% file: 2_background.tex
\section{Theory Development}

To theorize about the effects of license choice on a project's performance,
several strands of prior research may be relevant.
First is the line of work on motivations of
individuals~\cite{von2012carrots} and businesses~\cite{Harhoff02} to
participate in OSS.
Second, technical, social, and other factors
that either constraint certain choices in software development or
make them more convenient~\cite{ma2020methodology}.
Third, the specific plans and objectives of the project, and how each license choice may advance and hinder these goals~\cite{sen2008determinants}.

\subsection{Motivations}

In an extensive literature review, \citet{von2012carrots} categorized OSS participant motivations
into ideology, altruism, kinship, fun, reputation, reciprocity,
learning, own-use, career, and pay. 
Some of these motivations can probably
influence the type of licenses contributors prefer.
For instance, ideology and reciprocity
may favor strongly copyleft licenses (over all other
considerations), which require code sharing, while altruism may lean
toward more permissive licenses.
Kinship should drive adoption of
licenses used by developers' social network, while own-use
motivation appears to be orthogonal with license choice and may
allow for other factors, like convenience and compatibility, to 
dominate.

In case of commercial involvement, motivations may diverge from
individual contributors, even if they generally align with profit
maximization~\cite{Harhoff02}.
Studies of open and gated software
communities highlight the difficulties firms face in balancing
community-based value creation with private value
appropriation~\cite{Shah06}.
However, community sponsorship and
licensing strategies can address these tensions.
For instance,
restrictive licenses can ensure long-term control over
contributions, whereas nonmarket\footnote{
The term ``nonmarket'' excludes for-profit organizations.}
sponsors might alleviate
concerns about the project’s sustainability without the need for
restrictive licenses~\cite{stewart2006impacts}.
This highlights how
the nature of the sponsor—market or nonmarket—may influence
licensing strategies and developer motivations.

Empirical studies further emphasize the connection between license
choice and project structure. 
\citet{fershtman2007open} applied economic theories of motivation 
and found that factors such as status, 
signaling, or intrinsic motivation for participation in OSS projects are 
linked to restrictive licenses.
They observed that restrictive licenses 
attract more contributors per project, while permissive licenses result in 
higher productivity per contributor.
Similarly, research
into firm involvement in OSS shows that licensing choices may be
driven by strategic goals.
For instance, firms may adopt permissive
licenses to facilitate bundling proprietary and open source
code~\cite{economicos02}.
\citet{dahlander2008firms} identified three strategies
firms use to engage with OSS communities:
(1) accessing community development to extend resources, 
(2) aligning their strategy with community efforts, and 
(3) assimilating the community to integrate
and share outcomes.
Different commercial models necessitate distinct
licensing strategies.
\citet{wagstrom2010impact} and \citet{zhou2016inflow} identified 
two types of firm engagement:
community-focused firms (e.g., GNOME), which prioritize
building vibrant OSS communities and monetize through services
(which favor restrictive licenses so that a competitor can not
create proprietary enhancements without sharing the code), and
product-focused firms (e.g., Eclipse), which rely on product
revenues and thus favors weaker copyleft provisions to allow some
forms of bundling with proprietary code. 

Emergence of cloud services
endangered the business model of OSS companies that rely on service
contracts.
Because cloud companies provide computing services, they
can also run (and sell) the same OSS software as a service, thus
cutting out companies that developed the software.
For example, in 2021, Elasticsearch changed its Apache license to a
commercial one where the products built
from that code can not be provided to others as a managed service.
They made another change in 2024, moving to Affero GPL,
which is a copyleft license that requires anyone running a modified
program on a server and letting other users communicate with it
to also provide access to the source code of the modified 
version running on the server. 
As we can see, the more permissive license (Apache) was changed to a 
more restrictive one (commercial, then AGPL), as is typical for the service model.

Despite significant amount of research on the motivations to 
participate in OSS and commercial
involvement, few of the key predictors can be reliably
operationalized at scale.
Based on the \citet{fershtman2007open} 
conjecture that restrictive licenses attract more contributors per project,
we can expect that adopting a restrictive license will result in higher
contributor counts:
\begin{itemize}
    \item \textbf{Hypothesis (H1a)}: More restrictive licenses results in higher total numbers of authors.  
    \item \textbf{Hypothesis (H1b)}: More restrictive licenses (often ideologically motivated) are more likely to be retained.
\end{itemize}

\subsection{Social and Technical Choice}

Licenses, like other technologies and practices, spread through
communities.
Prior studies, such as those by \citet{ma2020methodology} and 
others~\cite{lamba2020heard, choice23}, used the
theory of social contagion~\cite{shoroye2015exploring} to estimate
the impact of exposure, susceptibility, and infectiousness on
developer choices regarding package (and tool) selection.
These studies found
that exposure measured by the number of overall deployments at the
time of choice, had a significant positive effect on 
selection.

We expect a similar pattern for license selection. One dimension of
exposure is the prevalence of a license (which is relatively
constant over time) and favors the selection of popular
licenses. i.e., developers who are indifferent or unaware of the
differences between licenses are likely to gravitate toward those
they have encountered.
Just as developers tend to choose packages
that are technically compatible with their existing technology stack
and socially aligned with their collaborators' choices, we
anticipate a similar behavior in license choice. Developers are
likely to select a license compatible with their current projects or
one already used by their collaborators (also, see ``Kinship''
motivation above).

The choice of programming language also influences license selection due to the culture and practices embedded within specific developer communities.
For example, languages like Python and JavaScript, commonly used in web development, often favor permissive licenses like MIT or Apache, which emphasize reuse and flexibility~\cite{lerner2005economics}.
In contrast, languages such as C or C++, which are dominant in systems programming, tend to be linked with copyleft licenses like GPL, prioritizing keeping derivative works open~\cite{fitzgerald2006transformation}.
The language captures the influence of community norms and technical compatibility on license selection.

The community size associated with a project can also impact license decisions.
\citet{tsay2014influence} found that projects with a large number of forks are perceived as more valuable and trustworthy by the community.
Projects aiming to encourage collaboration and reuse to maintain this momentum may favor permissive licenses.
Similarly, \citet{borges2018s} suggest that the number of stars signals community approval, which, like forks, may influence the choice of more permissive licenses.

Furthermore, the number of upstream/downstream projects a project interacts with
can play a crucial role in license selection.
If a project interacts with many downstream projects, it may opt for licenses that balance flexibility and control,
favoring weaker or conditional licenses over highly permissive or restrictive ones.

The decision to count upstream/downstream projects is motivated by the need to quantify a project's external interactions.
While analyzing the specific licenses of these projects would provide more 
detailed insights, counting them offers a practical approach to estimate the 
complexity and potential challenges related to license alignment. 
Based on this discussion, we hypothesize that:
\begin{itemize}
    \item \textbf{Hypothesis (H2a)}: The license choice is affected by overall popularity of the license 
    at adoption time.
    \item \textbf{Hypothesis (H2b)}: The license choice is affected by programming language culture-specific norms.
    \item \textbf{Hypothesis (H2c)}: More permissive licenses result in higher community size 
    (i.e., upstream and downstream projects).
\end{itemize}

\subsection{Project Goals}

While legal compliance sets essential boundaries, the specific needs and objectives of a project further influence the selection of an appropriate license.
Developers often choose licenses based on their preferences for the future use of the software. 
Research by \citet{sen2008determinants}
highlights that developers' motivations -- such as a desire for
widespread adoption or strong copyleft enforcement -- also play a significant role in license selection.

The choice of license can significantly influence a project’s development trajectory.
\citet{colazo2009impact} found that copyleft licenses promote greater permanence and long-term involvement in projects, affecting both developer contributions and engagement.

The number of commits is a key indicator of project activity and maintenance. 
\citet{koch2002effort} suggests that projects with frequent commits tend to have more active development cycles, making them more reliable and attractive for reuse.
As project activity increases, maintainers may select licenses that either encourage contributions or protect the evolving codebase from potential misuse, balancing the needs for stability and control.

The number of files in a project reflects its complexity and modularity.
\citet{mockus2007large} observed that larger projects with more files are more likely to offer reusable components, making license choice critical for balancing reuse and intellectual property protection.
\citet{bird2009putting} further suggested that projects with more files tend to have modular structures, which encourage community engagement and adoption, potentially favoring more restrictive licenses.

Time-based metrics -- such as the earliest and latest commits and the number of active
months -- provide insights into a project’s maturity and stability.
\citet{capiluppi2003characteristics} noted that older, more mature projects are often seen as more reliable, which could be influenced by the choice of licenses that support long-term sustainability, such as copyleft licenses.
On the other hand, newer projects may opt for permissive licenses to facilitate rapid contributions during active development phases.

Burstiness, as a proxy for fluctuating activity, was operationalized by measuring the number of months in which the project had activity over its lifetime (from earliest to latest commit).
Projects with high burstiness may not focus heavily on legal concerns, and instead tend to choose well-known, widely used licenses: 

\begin{itemize}
\item \textbf{Hypothesis (H3a)}: The license choice will affect the project activity approximated by number of commits.
\item \textbf{Hypothesis (H3b)}: More complex projects (higher number of files and blobs) will favor restrictive licenses.
\item \textbf{Hypothesis (H3c)}: Restrictive licenses support long-term sustainability as approximated by project activity duration.
\item \textbf{Hypothesis (H3d)}: Burstiness reflects the lack of
stability and favors more permissive licenses.
\end{itemize}

\section{Comparisons to Prior Work}

Despite legal importance of license choice and the potential to stunt
innovation and poison supply chains, it appears that a significant
portion of OSS participants are ignoring the need to choose a license
or to verify license compatibility. 
We construct a preliminary
theory of how license choice might affect the project by reviewing relevant
literature, operationalizing potential affected factors,
and testing that theory. 
However, while our work is not the first to empirically study software licenses,
it differs from the literature in two important ways.

\noindent \textit{Comprehensive Identification of Licenses.}
Most prior research, such as that by \citet{wu2024large} and \citet{xu2023lidetector}, primarily depends on explicit license declarations found in metadata files.
Others, like \citet{feng2019open}, apply static analysis on binaries to detect embedded license texts.
However, these methods may overlook licenses that are not clearly stated or are stored in unconventional directories.
In contrast, we use a comprehensive dataset by \citet{jahanshahi2024license}, 
compiled by scanning virtually the entire OSS ecosystem for any files containing ``license'' in their filepath.
Moreover, the authors use the winnowing algorithm, a robust method for matching license texts to known licenses, enhancing the accuracy of detecting both partial and full matches, even when the text is embedded or slightly modified.
This method captures not only standard license files but also other files potentially holding licensing details, ensuring that no relevant license information is missed.

\noindent \textit{Scale and Scope of Analysis.} 
Previous works often limit their scope to specific platforms (e.g., GitHub), a few package manager environments (e.g., NPM), or types of licenses (e.g., OSI-approved licenses), lacking a comprehensive, large-scale approach to detecting and analyzing licenses.
Our study expands this scope by analyzing essentially the entire open source landscape, ensuring a more comprehensive cross-platform understanding of licensing practices.

In Table~\ref{tbl:litRev}, we compare our study's methodology, coverage, scope, and scale with the most recent comparable studies in the field.
\begin{table*}[t]
\centering 
\caption{Latest Studies in Open Source Software Licensing}
\label{tbl:litRev}
\resizebox{0.95\textwidth}{!}{
\begin{tabular}{llllr}
    \toprule
    & \textbf{License Detection} & \textbf{Coverage} & \textbf{Scope} & \textbf{Scale} (Projects) \\
    \midrule
    \cite{wu2024large} & Package Manager API & License Files & Maven, NPM, PyPI, RubyGems, Cargo & 3,474,778 \\
    \cite{cui2023empirical} & NLP/Extract License Terms & All Files & GitHub - >1000 Stars & 16,341 \\
    \cite{wolter2023open} & GitHub API+NLP/Extract License Terms & All Files & GitHub - Cataloged by OpenHub & 1,000 \\
    \cite{xu2023lidetector} & NLP/Extract License Terms & All Files & GitHub - Popular Python Projects & 1,846 \\
    \multicolumn{5}{c}{\dotfill} \\
    \textbf{This Study} & \textbf{NLP/Matching Known Licenses} & \textbf{License Files} & \textbf{Entire OSS} & \textbf{131,171,379} \\
    \bottomrule
\end{tabular}}
\end{table*}

%% file: 3_methodology.tex
\section{Methodology}\label{method}

To test our hypotheses, we adopt a two-step methodology.
First, we gather descriptive statistics to quantify the prevalence, retention, 
and temporal trends of different license types within publicly available repositories.
We then identify projects that switched from permissive to restrictive 
licenses (or vice versa) and measure their performance changes before and after 
the switch, using a multivariate multiple regression approach.

\subsection{License Types}

We group licenses based on their characteristics.
This grouping helps categorize and understand the different ways software can be distributed and modified.
These categories typically include permissive, copyleft, weak copyleft, and public domain/unlicensed code.
This classification is widely recognized in the field~\cite{kaminski2007open} and is supported by various scholarly sources.

1. \textit{Permissive Licenses}: These licenses, such as the MIT and BSD licenses, are known for their minimal restrictions on how the software can be used.
They allow software to be freely used, modified, and redistributed, even as part of proprietary software~\cite{kapitsaki2019modeling}.
This permissiveness promotes wider adoption and integration of the software in diverse projects, including commercial applications~\cite{kapitsaki2022help}.

2. \textit{Copyleft Licenses}: These licenses, like the GNU General Public License (GPL), require that any modified versions of the software must also be distributed under the same license.
This ensures that derivative works remain free and open, thus preserving the original freedoms granted by the license~\cite{d2007copyleft}.
This characteristic is essential for maintaining the open source nature of software, as it prevents proprietary modifications~\cite{laurent2012free}.

3. \textit{Weak Copyleft Licenses}: These licenses, such as the GNU Lesser General Public License (LGPL) and Mozilla Public License (MPL), strike a balance between permissive and strong copyleft licenses.
They allow linking with proprietary software under certain conditions, while modifications to the licensed components themselves must remain open~\cite{alamoudi2020open}.
This flexibility encourages the use of open source libraries in both open and closed source projects~\cite{gamalielsson2017licensing}.

4. \textit{Conditional Open Licenses}: These licenses include specific conditions that must be met for usage, such as attribution (CC-BY), non-commercial use, or share-alike requirements.
Creative Commons licenses, like CC-BY and CC-BY-SA, provide a framework for sharing while respecting the creator's conditions~\cite{de2009creative}.

5. \textit{Public Domain/Unlicensed}: Public domain and unlicensed software, including those using the Creative Commons Zero (CC0) license, are not restricted by copyright law.
The authors of such software waive all rights, allowing anyone to use, modify, and distribute the work without any conditions.
This category provides the maximum freedom for software use and is often utilized for simple, non-critical software components.

Given 597 distinct licenses, we manually classified the top 50, covering 98.92\% of projects. 
The remaining ~1\% were grouped as “other” to maintain reproducibility and focus on 
licenses most relevant at scale.

\subsection{World of Code Infrastructure}

World of Code (WoC)\footnote{\url{https://worldofcode.org}} is an infrastructure designed to cross-reference source code change data across the entire FLOSS community, facilitating sampling, measurement, and analysis within and across software ecosystems~\cite{ma2019world,ma2021world}.
In essence, it is a software analysis pipeline that encompasses the discovery and retrieval of data, storage and updates, and the transformations and data augmentation required for downstream analytic tasks~\cite{ma2021world}.

WoC offers various maps that connect git objects and metadata (commits, blobs, authors) to each other.
It also provides higher-level maps, such as project-to-data connections (e.g., project-to-author), author aliasing~\cite{fry2020dataset}, and project deforking maps~\cite{mockus2020complete}. 
We use the project-to-license (P2L) map~\cite{jahanshahi2024license} from WoC that provides all the times at which a license was committed to a project and also verifies whether it still exists in the project's latest version.\footnote{
Version V, latest at the time of this study.}
We also employ the concept of deforked projects as introduced by~\cite{mockus2020complete} to avoid potential biases from forks and duplicates of the same project.
Throughout this paper, the term ``project'' refers to this deforked project unless stated otherwise.

\subsection{License Change}

We use license changes within public repositories to test our hypotheses 
on how license choice affects project performance. A project's decision 
to switch licenses should constitute a highly deliberate, informed event, 
typically involving careful consideration of implications related to legal 
constraints, developer participation, and user adoption. Unlike
initial license selections -- which may be arbitrary or uninformed -- license
changes reflect strategic intent, often in response to evolving community
needs, competitive pressures, or growth ambitions. This intentionality
makes license changes ideal natural experiments, enabling clearer,
quasi-causal inferences regarding the impact of licensing decisions on 
measurable performance metrics. By focusing specifically on projects 
that have altered their licenses, this study leverages within-project 
variation, inherently controlling for stable project characteristics, 
thereby significantly strengthening internal validity and enhancing 
the precision of conclusions drawn from the analysis.

For example, moving to a more restrictive license type might be chosen to 
protect the project's intellectual property or to enforce open-source
principles more rigorously.
\citet{kechagia2010open} illustrate how different license types affect 
dependency management and compliance in software projects.

To investigate the impact of license switching, we first identify projects
with more than one license throughout their lifetime.
We then refine this group to those that adopted exactly one license at their 
initial license adoption time and maintained exactly one license at their latest
recorded status, thereby excluding any projects that held multiple licenses 
simultaneously at either point in time.

\begin{figure*}[t]
\centering
    \includegraphics[width=0.8\textwidth, clip=true, trim=0 30 0 0]{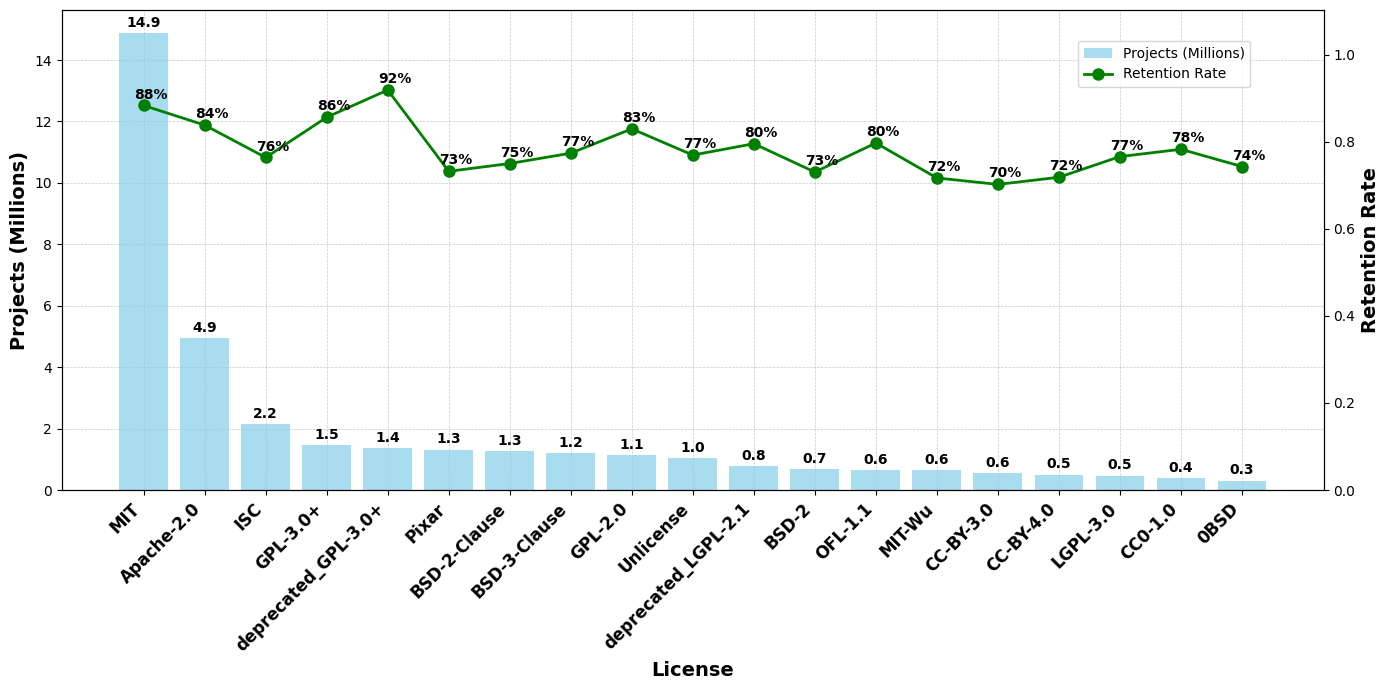}
    \caption{The distribution and retention rates of the most used licenses.}
    \label{fig:dis1}
    \Description{The distribution and retention rates of the most used licenses.}
\end{figure*}

This also means that if a project starts with license A, then changes to B,
and later changes to license C, we consider it as one change from A to C, 
rather than two separate changes.
Consequently, we excluded projects where the first license type and 
the latest license type were the same.
Although these projects had other licenses at some point, they were 
removed, resulting in no change between the first and last license types.

Next, we categorize licenses into two high-level groups: ``permissive'', which 
includes public-domain and permissive licenses, and ``restrictive'', encompassing 
copyleft, weak-copyleft, and conditional open licenses.
Within these groups, we select projects that switched from a permissive to a 
restrictive license or vice versa.

To measure performance changes, we compute key project metrics during two separate 
one-year intervals: one immediately following the first license adoption and 
another immediately following the last license adoption.
Projects are excluded if less than one year separates the adoption dates of the
first and last licenses or if less than one year has elapsed since the last 
license adoption at the time of our data curation. 
We then model the change in these metrics (i.e., metric values in the post-switch 
interval minus those in the pre-switch interval) based on the direction of the 
license change (restrictive to permissive or permissive to restrictive).
We control for the project’s primary language, its start and 
end times, the delay before the first license adoption, the elapsed time between
the first and last license adoptions, and the proportion of new OSS projects 
adopting that license type at the time of the last license change.

To test our hypotheses, we selected a regression model appropriate to the 
nature of our response variable, specifically, multivariate multiple regression.
This model is an extension of multiple 
regression that allows for multiple dependent variables (outcomes) to be predicted
by multiple independent variables (predictors) simultaneously.
This model is useful when the dependent variables are correlated and need to be 
analyzed together to improve prediction accuracy and interpretation.

%% file: 4_results.tex
\section{Results and Discussion}

We organize our results in three parts. 
First, we present a broad exploration of different license types, highlighting 
their prevalence, retention, and general trends.
Next, we detail the regression analysis that examines how the choice of license 
relates to project performance. 
Finally, we focus on the phenomenon of license switching by analyzing performance
metrics before and after a change from permissive to restrictive (or vice versa).

\subsection{Types and Prevalence of Licenses}

\subsubsection{Our Findings}

We first count how many projects had at least one occurrence 
of any license during their lifetime,
then identify the top 20 most used licenses, 
as shown in Figure~\ref{fig:dis1}, and analyze their frequency.
Out of the 131,171,379 projects indexed by WoC, 22,281,342 projects had at 
least one license committed to their project at some point in their lifetime.
However, when examining the latest version of these projects, 
this number drops to 20,110,256 projects.
When interpreting the numbers in the figure, note 
that the sum of per-license counts exceeds the total number of projects with a license.
This discrepancy arises because some projects have multiple licenses and 
are therefore counted in several categories.

To calculate the license retention rates, we count how many 
projects still had the license in their latest state and divide this number 
by the total number of projects that had ever used the license.
This retention rate, also in Figure~\ref{fig:dis1}, indicates the proportion of projects that continue to use the license over time.

As shown in the figure, the MIT license is the most prominent, followed
by the Apache 2.0 and ISC licenses.
Notably, per H1b, the deprecated GPL 3 license (with strong
copyleft provisions) has the highest retention rate.
In contrast, the Creative Commons 3 license (CC-BY-3.0) 
has the lowest retention rate.

\begin{figure}[t]
\centering
    \includegraphics[width=\linewidth, clip=true, trim=0 25 0 0]{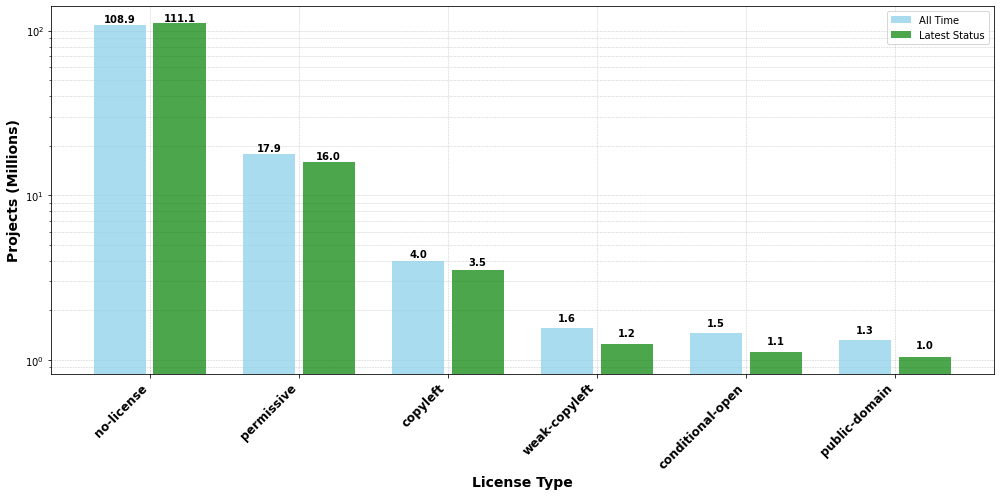}
    \caption{The distribution of license types across projects.}
    \label{fig:dis3}
    \Description{The distribution of license types across projects.}
\end{figure}

We also analyze the distribution of projects according to their license types, in two conditions.
First, we consider the distribution across the entire project history, 
meaning that if a project adopted a license at any point and later removed it, 
it is still counted as having that license.
Second, we determine the distribution based on the most recent status of 
each project.
The results in Figure~\ref{fig:dis3} show that ``conditional-open'', ``weak-copyleft'', 
and ``public-domain'' license types have the lowest
retention, with approximately a quarter of the projects that ever had 
such license changing it by their latest version.
For comparison, only approximately 8\% of ``copyleft'' licenses were changed, 
which is consistent with our theory (H1b) that ideology-related license choice should be most ``sticky''.

\begin{figure}[t]
    \centering
    \includegraphics[width=\linewidth, clip=true, trim=0 5 0 0]{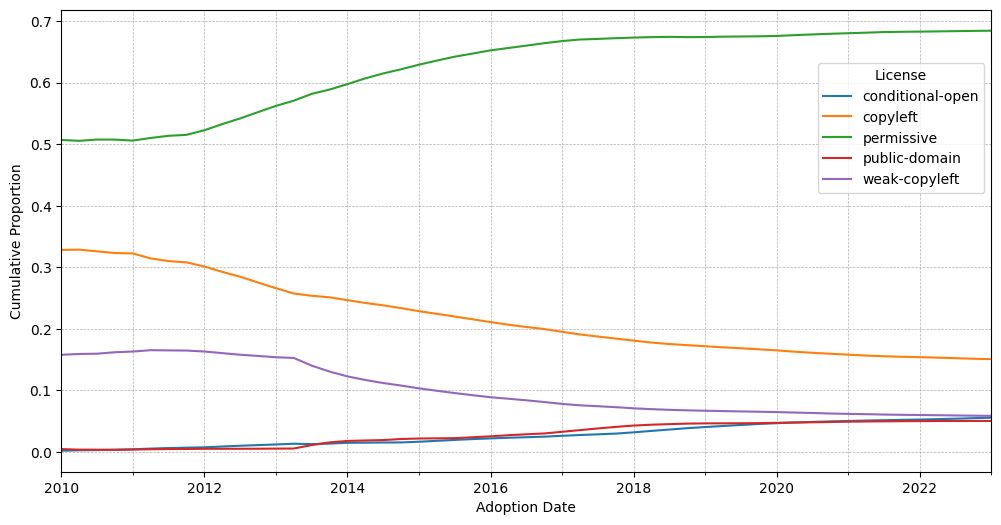}
    \caption{Proportion of license type over time}
    \label{fig:proportion}
    \Description{Proportion of license type over time}
\end{figure}

We also analyze the proportion of adopted licenses at each point in time, 
categorized by license type, to identify trends in licensing preferences over 
time. This allows us to observe shifts in the adoption of different 
licenses and assess whether certain license types have gained or declined 
in popularity. By tracking these proportions longitudinally, we can better 
understand how licensing choices evolve and whether external factors, such 
as regulatory changes or community norms, influence these trends.
The results are shown in Figure~\ref{fig:proportion}.
Looking at the trends we clearly see that although copyleft licenses
may be the most sticky, the tendency to adopt these licenses has decreased
over time (H2a).

\subsubsection{Comparison with Literature}

Table~\ref{tab:res1a} provides a comparison between the results in the 
literature and our study, highlighting substantial differences primarily 
due to variations in scope, methodology, and dataset size 
(see Table~\ref{tbl:litRev}).

A key methodological difference between our study and \citet{wu2024large} lies in the treatment of data points.
\citet{wu2024large} considered each version of a package as a separate data point, 
resulting in 46.59 million data points from 3.47 million projects.
This approach could overrepresent projects with many
versions, potentially skewing adoption trends.
Our study treats each project as a single data point, providing 
a more balanced view of the ecosystem without overemphasizing 
projects with frequent updates.

The focus of the \citet{wu2024large} study on only five package managers might explain 
the significant discrepancy in the percentage of projects without a license:
21.12\%
in their case,\footnote{
Since the paper provides these percentages by package manager, we calculated 
a weighted average based on the number of 
projects per package manager.} 
versus 83.01\% in ours.
On the other hand, \citet{cui2023empirical} reported that only 10.51\% of
projects were unlicensed.
The difference with \citet{wu2024large} likely stems from the fact that the
package managers they examined are associated with more mature and 
well-maintained projects, which are more likely to have defined licenses.
Additionally, \citet{cui2023empirical} filtered their data to include 
only projects with over 1,000 stars, which might explain why the 
percentage of unlicensed projects is so low.\footnote{
These could also be influenced by the differences between OSS projects
and publicly available projects as was explained in the introduction.}
In contrast, our dataset is broader, including projects which may be less mature or less formally governed, leading to a higher incidence of unlicensed projects.

\begin{table*}[t]
\centering
\caption{Comparison between our results and the literature.}
\label{tab:res1a}
\begin{tabular}{lrr|rrrrr}
    \toprule
    & \textbf{Projects} & \textbf{No License} & \textbf{Copyleft} & \textbf{Weak} & \textbf{Conditional} & \textbf{Permissive} & \textbf{Public} \\
    & & & & \textbf{Copyleft} & \textbf{Open} & & \textbf{Domain} \\
    \midrule
    \cite{wu2024large} & 3,474,778 & 21.12\% & 5.72\% & 5.12\% & - & 89.16\% & - \\
    \cite{cui2023empirical} & 16,341 & 10.51\% & 11.32\% & 0.97\% & - & 87.71\% & - \\
    \textbf{This Study} & 131,171,379 & 83.01\% & 15.35\% & 5.26\% & 4.82\% & 70.18\% & 4.39\% \\
    \bottomrule
\end{tabular}
\end{table*}

The analysis of copyleft versus permissive licenses further reflects 
the differences in scope.
\citet{wu2024large} found a preponderance of permissive licenses, with much 
lower percentages for copyleft and weak copyleft licenses.
However, our study, covering all programming languages, reveals a more 
diverse licensing landscape, with a lower percentage of permissive licenses
and higher percentages of copyleft licenses.
This suggests that by focusing on just five package managers, \citet{wu2024large}
might underrepresent broader licensing trends, particularly the prevalence 
of copyleft licenses across the global open source ecosystem.

\begin{figure}[h]
\begin{tcolorbox}[colback=gray!5!white, colframe=gray!70!black, title=Key Findings]
\small
\begin{enumerate}[leftmargin=4mm]
    \item Overall, 83\% of the projects in the dataset lack a formal license, which is significantly higher than previously reported.
    \item The MIT license is the most widely used license,
      accounting for 65\% of all licensed projects and 10\% of all
      public projects.
    \item Despite the overall preference for permissive licenses,
      copyleft licenses such as GPL have high retention rates
      cf.\ H1b.
    \item The differences in license usage reported in our study compared 
    to prior work highlight how sensitive such analyses are to sampling strategies.
\end{enumerate}
\end{tcolorbox}
\Description{Key Findings}
\end{figure}

\subsubsection{Implications}

The analysis of public repository licensing reveals that 83\% of projects lack a
formal license, posing significant legal risks and discouraging
innovation by preventing responsible actors from reusing unlicensed code. 
This highlights the need for developers to prioritize licensing to set 
clear terms of use.
Our findings challenge previous findings, which underestimate the 
prevalence of unlicensed projects by focusing on mature ones, missing 
the broader risks of software reuse.

Even though many of these unlicensed projects may seem trivial and 
technically not OSS, their 
public availability still creates legal ambiguity and poses a risk 
of license noncompliance, especially when combined with the lack of 
thorough legal review in many downstream projects.
\citet{jahanshahi2024beyond} demonstrated that in the context of 
copy-based reuse, nearly 18\% of reused artifacts originated from 
very small projects,\footnote{
Projects with no stars and fewer than 10 commits.
} while large projects\footnote{
Projects with more than 10 stars and 100 commits.
} accounted for only 32\% of reused artifacts.
This highlights the importance of addressing the issue of unlicensed code within the OSS ecosystem.

Among projects that are licensed, the MIT license is the most prevalent, 
reflecting a historic preference for permissive licenses that facilitate 
broad adoption and collaboration.
However, the continued use of copyleft licenses like the GPL
(which was popular decades ago)
demonstrates that many developers are committed to preserving the
open nature of derivative works.

\subsection{The Impact of License Changes}

\begin{figure}[t]
\centering
\includegraphics[width=0.8\linewidth, clip=true, trim=10 0 90 0]{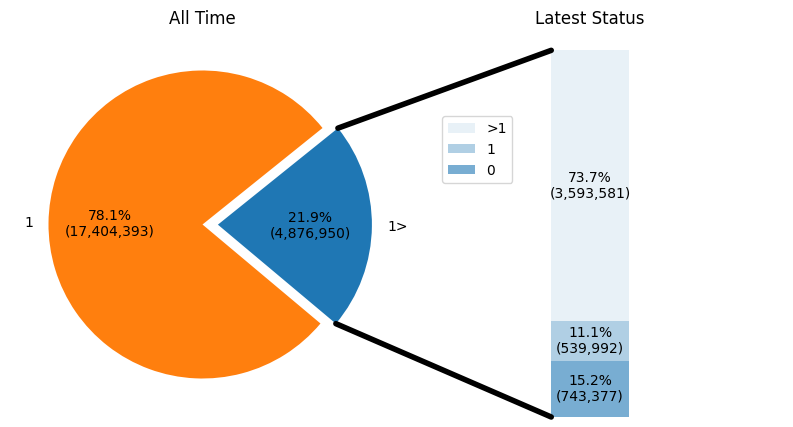}
\caption{Distribution of license type count in projects}
\Description{Distribution of license type count in projects}
\label{fig:pie}
\end{figure}

Examining the distribution of license type counts in projects 
(Figure~\ref{fig:pie}), we observe that nearly 22\% of projects had more 
than one license type in their lifetime.
Among these,
74\% still had more than one license type in their latest status as well, indicating they did not 
necessarily change their license.
On the other hand, 15\% of these projects do not have any license in their 
latest status, meaning they deleted their license.
The middle group, consisting of projects that had more than one license type 
in their lifetime but have only one type ultimately, is the group
we analyze license changes in.

\subsubsection{Our Findings}

Now we turn to the results of our regression model discussed in 
the Methodology section.
The basic statistics of the response and control variables, including the 
5th percentile, median, mean, and 95th percentile for the numeric variables, 
as well as the counts of different levels for the factor variables, 
are presented in Table~\ref{tbl:stat-l}.

\begin{table*}[t]
\centering
\caption{License change model - descriptive statistics.}
\resizebox{0.95\textwidth}{!}{
\begin{tabular}{cllcccc}
  \toprule
  & \textbf{Variable} & \textbf{Description} & & \textbf{Statistics} & & \\ 
  \midrule
  \midrule
  \multirow{9}{*}{\rotatebox{90}{\textbf{Response}}} 
  & & & \textbf{5\%} & \textbf{Median} & \textbf{Mean} & \textbf{95\%} \\ 
  & AuthorsDiff & Difference in number of authors & -5 & 0 & 0.91 & 6 \\
  & BlobsDiff & Difference in number of blobs & -1,470 & -17 & 56.6 & 1,059.2 \\
  & CommitsDiff & Difference in number of commits & -309.1 & -6 & 0.83 & 209 \\
  & FilesDiff & Difference in number of files & -856.1 & -4 & 116.92 & 634 \\
  & ActiveMonDiff & Difference in number of active months & -8 & -1 & -1.06 & 5 \\
  & UpProjectsDiff & Difference in number of upstream projects & -35 & -1 & -3.54 & 16 \\
  & DownProjectsDiff & Difference in number of downstream projects & -7 & 0 & 19.68 & 25.1 \\ 
  & BurstinessDiff & Difference in burstiness (duration / active months) & -7 & 0.28 & 1.27 & 10.5 \\ 
  \midrule
  \multirow{8}{*}{\rotatebox{90}{\textbf{Control}}} 
  & EarliestCommit & Project age (months) & 39.87 & 84 & 90.05 & 162.04 \\
  & LatestCommit & Project latest inactivity (months) & 3.23 & 27.57 & 34.95 & 87.64 \\
  & AdoptDelay & Earliest commit to license adoption (months) & 0 & 0.13 & 3.35 & 19.37 \\ 
  & Distance & Distance between first and last adoption (months) & 12 & 24 & 30.77 & 73 \\
  & Proportion & Proportion of projects adopting same license at change time & 0.05 & 0.64 & 0.44 & 0.68 \\
  \multicolumn{6}{c}{\dotfill} \\
  
  & Language & \hspace{-0.2cm}JavaScript \hspace{1cm} Python \hspace{1.15cm} C/C++ \hspace{1.3cm} Java & PHP & Go & Ruby & Other \\ 
  
  & Counts (\%) & \hspace{-0.4cm}7,566 (32.03\%) \hspace{0.3cm} 3,708 (15.70\%) \hspace{0.3cm} 3,027 (12.82\%) \hspace{0.3cm} 2,371 (10.04\%) & 1,782 (7.54\%) & 526 (2.23\%) & 518 (2.19\%) & 4,121 (17.45\%) \\ 
  \midrule
  & Change & \hspace{-1.65cm} Direction \hspace{3cm} Restrictive to Permissive & Permissive & \hspace{-2cm} to & \hspace{-3.75cm} Restrictive \\ 

  & Counts (\%) & \hspace{3.3cm} 14,329 (60.67\%) & & \hspace{-3cm} 9,290	& \hspace{-4.6cm}(39.33\%) \\ 
  
  \bottomrule
\end{tabular}}
\label{tbl:stat-l}
\end{table*}

Given that we have multiple response variables, the model provides a 
distinct set of coefficients for each outcome.
Each set of coefficients indicates the influence of the predictor variables 
on the probability of the response variable assuming that there might 
be correlation between response variables.

For language variable, we employed sum contrasts, also known as effect 
coding, where each level of the predictor variable is compared to the 
overall mean of all levels.
This approach is particularly advantageous in models where the goal is to 
compare each category to the overall mean rather than to a specific 
reference category, as it offers a more symmetric interpretation of the effects.
In the sum contrast method, the sum of the coefficients for all levels,
including the intercept, must equal zero.
We also include the interaction term between language and license 
change in the model.

The change direction variable has two levels, with Permissive to Restrictive 
(P2R) set as the base level. As a result, the model outputs only display 
coefficients for Restrictive to Permissive (R2P), which indicate the 
difference relative to the base level (P2R).

Furthermore, when interpreting the earliest and latest commit times, 
it is important to note that the predictors in the model represent the 
time elapsed since those commits. A higher earliest commit value indicates
that the commit occurred further in the past, meaning the project is older, 
whereas a lower value suggests a more recent commit. Similarly, a higher 
latest commit value means the project has been inactive for a longer period.

Table~\ref{tbl:anova-lc} shows Type III MANOVA Tests with Pillai test statistic
for the fitted model. 
The results show that the p-values for all the predictors are close to zero, 
indicating that each predictor significantly improves our model's fit.

\begin{table}[t]
\centering \small
\caption{License change model - Type III MANOVA Tests: Pillai test statistic}
\begin{tabular}{lccc}
    \toprule
    \textbf{Variable} & \textbf{DF} & \textbf{Test Stat} & \textbf{p.value} \\ 
    \midrule
    (Intercept) & 1 & 0.024183 & \cellcolor{lightgreen}$<2\times10^{-16}$ \\
    Change Direction & 1 & 0.002403 & \cellcolor{lightgreen}$1.99\times10^{-9}$ \\
    Language & 7 & 0.012655 & \cellcolor{lightgreen}$<2\times10^{-16}$ \\
    EarliestCommit & 1 & 0.061345 & \cellcolor{lightgreen}$<2\times10^{-16}$ \\
    LatestCommit & 1 & 0.111041 & \cellcolor{lightgreen}$<2\times10^{-16}$ \\
    Delay & 1 & 0.028073 & \cellcolor{lightgreen}$<2\times10^{-16}$ \\
    Distance & 1 & 0.018359 & \cellcolor{lightgreen}$<2\times10^{-16}$ \\
    Proportion & 1 & 0.001388 & \cellcolor{lightgreen}$6.79\times10^{-5}$ \\
    Change:Language & 7 & 0.008897 & \cellcolor{lightgreen}$<2\times10^{-16}$ \\
    \bottomrule
\end{tabular}
\label{tbl:anova-lc}
\end{table}

Model fit was evaluated using R² from univariate regressions, ranging from 0.02 to 0.09 
across outcomes-modest but typical for large-scale socio-technical data with high variability. 
Residuals showed no irregular patterns. Multicollinearity was assessed via generalized 
variance inflation factors (GVIFs); all predictors had adjusted $GVIF^{1/(2 \times Df)}$ 
values below standard thresholds, except the license change variable ($3.22$), which 
remains within acceptable limits given the model’s interactions and categorical terms.

\begin{figure*}[t]
    \centering
    \includegraphics[width=0.9\linewidth, clip=true, trim=0 18 0 0]{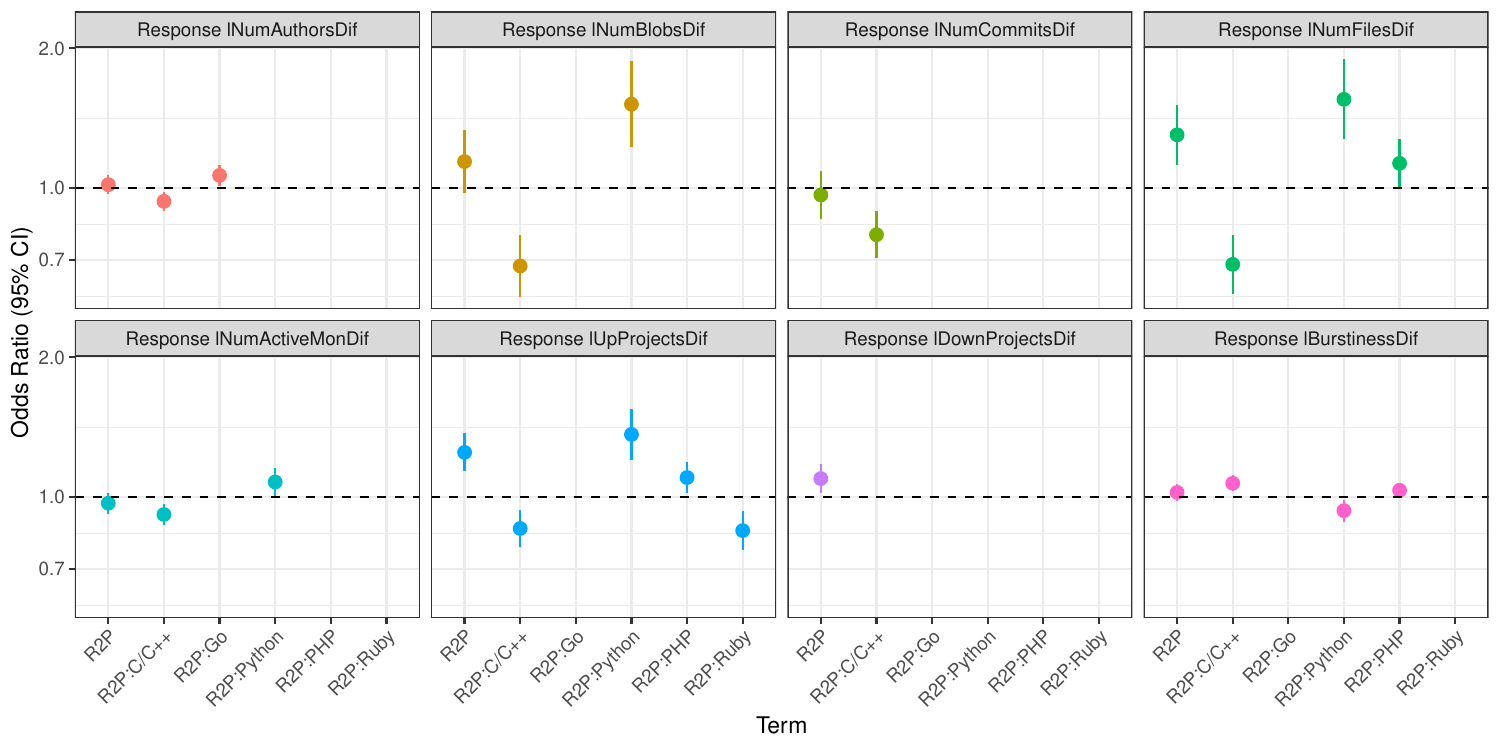}
    \caption{License change model - odds ratios}
    \label{fig:odds}
    \Description{License change model - odds ratios}
\end{figure*}

To better understand the effect of license change direction on project metrics,
we calculate the odds ratios for the change direction predictor and its 
interaction with language, considering only those predictors that are 
statistically significant at the 95\% confidence level. This ensures that 
the results focus on meaningful relationships rather than noise. 
Figure~\ref{fig:odds} displays these significant odds ratios along with 
their 95\% confidence 
intervals.\footnote{Please refer to the replication package for the full 
set of coefficients and p-values.} 
Since the analysis is based on a multivariate regression model,
the odds ratios are presented across different output variables, allowing
for a comparison of how the change direction predictor and 
its interaction with language influence various aspects of the model.

In these models, each coefficient represents a change in the 
log-odds (or log-count) of the outcome. The main R2P effect 
\(\beta_{\text{R2P}}\) 
shows how the outcome shifts when moving from a restrictive to 
a permissive license under sum contrasts (i.e., averaged across 
language levels). A significant R2P:Language interaction 
\(\beta_{\text{R2P:Lang}}\) 
adds to (or subtracts from) the overall R2P coefficient for 
that specific language. To obtain the total 
restrictive\(\rightarrow\)permissive effect for a given language, 
we sum \(\beta_{\text{R2P}} + \beta_{\text{R2P:Lang}}\) 
and then exponentiate. The result is the odds ratio -- a multiplier on 
the odds of an increase in the outcome metric compared to the 
permissive\(\rightarrow\)restrictive direction.

For the difference in the number of authors, C/C++ exhibits 
an odds ratio of about 0.95, suggesting these projects are roughly 5\% 
less likely to see increased authorship after moving from 
a restrictive to a permissive license compared to the opposite 
shift. In contrast, Go shows an odds ratio of about 1.08, 
indicating an 8\% greater likelihood of authorship growth
under permissive licensing. Although these effects are relatively 
modest, they demonstrate clear language-specific differences in
how communities respond to changes in license direction. While this partially
confirms H1a for C/C++ projects, it is not the same for all languages.

Regarding the difference in the number of blobs, C/C++ projects 
exhibit an odds ratio of about 0.78, indicating these projects 
are roughly 22\% less likely to experience blob growth when shifting 
from restrictive to permissive licenses (compared to going
permissive to restrictive). By contrast, Python shows an odds 
ratio of about 1.73, suggesting a 73\% higher likelihood of 
increased repository blob counts under permissive licensing. 
The notable magnitude of these odds ratios emphasizes that 
licensing direction can strongly shape development dynamics 
in these languages.

For the difference in the number of files, the overall restrictive to
permissive license shift has an odds ratio of about 1.30, indicating
roughly 30\% higher odds of increased file counts. However, this 
positive effect is notably reduced in C/C++, where the odds ratio is 
about 0.89 (around 11\% lower odds). In contrast, PHP and Python
have odds ratios near 1.47 and 2.02, respectively, suggesting that 
PHP projects are about 47\% more likely and Python projects are about 
100\% more likely to experience file-count growth under permissive licensing.

C/C++ (often complex projects) reduced growth (blobs/files) after shifting
to permissive licenses that suggests restrictive licenses were 
originally preferred to manage complexity effectively, partially supporting H3b.

In the difference in the number of commits, only C/C++ shows a statistically 
significant result, with an odds ratio of about 0.77—indicating a roughly 
23\% lower likelihood of increased commit frequency when shifting to a 
permissive license compared to the opposite direction. This moderate odds 
ratio suggests a tangible decrease in development activity within 
C/C++ communities following a move to permissive licensing, partially
supporting H3a.

Concerning the difference in active months, C/C++ projects show an odds 
ratio of about 0.89, indicating roughly 11\% lower odds of sustained
activity after moving from a restrictive to a permissive license,
partially supporting H3c.
By contrast, Python has an odds ratio of about 1.05, suggesting 
around 5\% higher odds of maintaining longer periods of active 
development under permissive licenses. Although these effects are modest, 
they illustrate clear language-specific variations.

Regarding the difference in upstream projects, the overall restrictive
to permissive shift shows an odds ratio of about 1.25, indicating roughly 
25\% higher odds of increased reuse from upstream. The interactions for
C/C++ and Ruby, though negative relative to the overall effect, still
place these languages slightly above 1 (about 1.07 and 1.06, respectively).
In contrast, PHP stands at about 1.38, and Python reaches about 1.70, 
suggesting considerably greater odds of upstream integration under 
permissive licensing. These variations across languages emphasize 
the nuanced ways in which license directionality can shape project
dependencies.

For the difference in downstream projects, the overall effect of 
going from a restrictive to a permissive license is an odds ratio of 
about 1.10, indicating roughly 10\% higher odds of increased downstream
usage or integration. This moderate boost suggests that 
adopting a permissive license tends to enhance a project’s
popularity within broader software ecosystems.

Lastly, for the difference in burstiness, the language-specific 
interactions point to moderate yet meaningful distinctions. C/C++ 
projects have an odds ratio of about 1.10 (around 10\% greater odds
of more intense activity bursts) under permissive licensing, while PHP 
stands at about 1.06 (6\% higher odds). By contrast, Python’s total
odds ratio is about 0.96, indicating roughly 4\% lower odds of
experiencing bursty activity. Although modest, these differences 
highlight notable variations in development dynamics associated 
with licensing direction. This shows partial support for H3d,
highlighting once again the nuances between different programming languages.

These quantified magnitudes, along with their directions, 
clearly demonstrate that license transitions influence project 
metrics meaningfully, with notable variations across different 
metrics and programming languages.

\begin{figure}[h]
\begin{tcolorbox}[colback=gray!5!white, colframe=gray!70!black, title=Key Findings]
\small
\begin{enumerate}[leftmargin=4mm]
    \item C/C++ projects see reduced authorship, commits, files, blobs, and activity after switching to permissive licenses cf.\ H2b.
    \item Python projects see increased repository growth and sustained activity under permissive licenses cf.\ H2b.
    \item  PHP projects benefit from permissive licenses through increased file additions and external project integration cf.\ H2b.
    \item Go projects attract moderately more authors after permissive license adoption cf.\ H2b.
    \item Permissive licenses generally increase both upstream and downstream project adoption. cf.\ H2c.
    \item Ruby projects show reduced upstream integration when adopting permissive licenses.
    \item License transitions have a moderate effect on burstiness: permissive licenses increase bursty activity for C/C++ and PHP, but reduce it for Python.
\end{enumerate}
\end{tcolorbox}
\Description{Key Findings}
\end{figure}

\subsubsection{Implications}

Programming language implies a broader choice of technology,
libraries, and tools. Thus, users of that entire ecosystem are probably nudged toward
the most common licenses used within it.
For example, C-based projects often prefer restrictive licenses, 
while Go and Python projects lean toward permissive ones.
Developers should be aware of how their language community influences 
licensing norms, and the community could offer language-specific
guidelines to align new projects with licensing best practices.

In the C/C++ 
community, where switching to permissive licenses slightly discourages 
new contributors, project maintainers might need to pair such license
transitions with additional strategies -- such as enhanced outreach or
clearly defined governance policies -- to retain or attract authors.
Conversely, in the Go community, the positive response to permissive
licenses indicates that maintainers could effectively use permissive
licensing as a practical tool to encourage contributor participation,
enhancing community building without significant additional effort.

Repository growth implications, measured by the number of blobs,
are particularly impactful. Python communities experience notable
growth under permissive licenses, indicating maintainers can 
strategically leverage permissive licensing to attract broader
contributions, resulting in richer functionality and greater innovation.
For C/C++, however, the practical advice would be cautious—adopting
permissive licenses without supplemental incentives might 
unintentionally discourage contributors from investing in 
substantial new content. Thus, maintainers in these communities 
should carefully weigh the trade-offs or introduce additional
motivators (e.g., clearer contributor recognition or rewards) 
alongside permissive licensing decisions.

Since permissive licensing tends to lower commit frequency in C/C++,
those community leaders considering
permissive licensing should prepare targeted interventions, such as
clearly articulated development roadmaps, contributor guidelines,
or engagement incentives to sustain development momentum after
a licensing change.

Implications for file-count differences indicate concrete strategic 
guidance. Given that PHP and Python projects substantially
benefit from permissive licenses regarding file creation,
maintainers in these ecosystems can confidently adopt permissive 
licenses when aiming for feature expansion or modularization. 
Conversely, C/C++ maintainers should be aware that permissive licenses
may require additional supporting strategies—such as enhanced
documentation or modular project structures—to ensure continued growth.

Practical considerations around project activity duration 
(active months) suggest nuanced strategies: Python projects benefit 
slightly in terms of sustained activity from permissive licenses, 
indicating that maintainers aiming for long-term, stable contributions 
can adopt permissive licenses confidently. In contrast, C/C++ maintainers
should recognize that permissive licensing may slightly shorten sustained
activity periods, making supplemental engagement strategies crucial -- such 
as periodic contribution drives or community recognition -- to 
sustain longer-term involvement.

Insights on upstream project usage have clear practical 
ramifications. Python and PHP communities'
increased usage of external OSS projects after adopting permissive
licenses suggests maintainers might strategically use permissive
licensing to encourage their projects' broader integration into
complex OSS ecosystems. Conversely, C/C++ and Ruby projects that 
see decreased usage of external dependencies under permissive licenses 
could benefit from using licensing strategically to foster self-contained,
tightly controlled project environments—useful in contexts requiring
high stability or predictable dependency management.

The moderate improvement in downstream project adoption following
permissive licensing highlights practical licensing guidance 
for maintainers seeking wider adoption. Project maintainers 
aiming for their software to become foundational or widely
used dependencies can practically adopt permissive licenses as
a direct strategy to facilitate their projects' broader use across
diverse OSS communities and applications.

Lastly, implications regarding burstiness—concentrated periods 
of intense development—offer actionable guidance. Since permissive
licensing boosts burst-like activity in C/C++ and PHP, maintainers
aiming for rapid or intensive development cycles (e.g., to meet release 
deadlines or achieve specific project milestones) could strategically
employ permissive licensing to trigger and sustain such heightened 
activity. Conversely, Python communities, experiencing steadier 
activity under permissive licenses, may practically leverage permissive
licensing for achieving predictable and consistent development flows 
rather than periodic intensive bursts.

Together, these practical insights enable OSS maintainers
to make informed licensing choices strategically tailored to specific
community dynamics, desired project outcomes, and broader ecosystem goals.

%% file: 5_limitations.tex
\section{Limitations}

The license map (P2L)~\cite{jahanshahi2024license} used for our study 
relies on detecting license files committed to a
project's repository. While this approach does not assume that
licenses are consistently and accurately recorded in dedicated
license.md files, it does not check the content of all
files. In practice, licenses might be
specified within individual source files, potentially leading to
underreporting or misclassification of a project's licensing status.

Additionally, while the P2L map captures license changes over time,
it may not fully account for licenses that were temporarily removed
or altered before being reinstated, which could introduce
inaccuracies in assessing a project's long-term licensing practices.
The reliance on the latest status of licenses might also obscure
important historical context, such as changes in licensing
throughout a project's lifecycle, which could be crucial for
understanding the project's evolution and the factors influencing
license choices.

Despite the efforts of manual verification employed in P2L map,
these limitations underscore the need for caution when interpreting
results based on it. Complementary methods, such as
cross-referencing with other data sources, may be necessary to
obtain a more complete and accurate picture of licensing practices
in open source software projects.

%% file: 7_conclusions.tex
\section{Conclusions}
The study offers a discussion of theoretical concepts combined with
rigorous statistical data analysis of the factors influencing
license choices in open source software projects.
Not all predictions were bourne out, suggesting that prior work used
to support the hypothesized effect was either no longer relevant or
obtained in a different context. Key finding that over 80\% of the
projects have no license making their code legally virtually unusable
highlight both risks of inadvertent poisoning of the software supply
chain and, perhaps more importantly, the obstacle for innovation
through its wider use. Many of the theorized predictions were
supported by the analysis, yet there was almost always meaningful 
differences between different programming languages, suggesting 
that a more mature theory of
license choice could be developed. The observed relationships may
help develop recommenders that can provide projects with the most
suitable default license and suggest changes as project matures or
its goals change.

%% file: 8_appendix.tex
\section{License Types}\label{ap:type}
List of SPDX license identifiers aggregated by their respective license types:

\textbf{Permissive}: 0BSD, AFL-3.0, Apache-2.0, BSD-2, BSD-2-Clause, BSD-3-Clause, BSL-1.0, ISC, Libpng, MIT, MIT-0, MITNFA, MIT-Wu, MS-PL, OpenSSL, PHP-3.01, Pixar, PSF-2.0, Ruby, SGI-B-2.0, TCL, WTFPL, Zlib

\textbf{Copyleft}: deprecated\_AGPL-3.0, deprecated\_GPL-3.0+, GPL-2.0, GPL-3.0+, GPL-CC-1.0, OSL-3.0

\textbf{Weak Copyleft}: Artistic-1.0-Perl, Artistic-2.0, CDDL-1.0, deprecated\_LGPL-2.1, eprecated\_LGPL-3.0, EPL-1.0, EPL-2.0, LGPL-2.0+, LGPL-3.0, MPL-1.1, MPL-2.0-no-copyleft-exception

\textbf{Conditional Open}: CC-BY-3.0, CC-BY-4.0, CC-BY-SA-3.0, CC-BY-SA-4.0, ODC-By-1.0, OFL-1.0, OFL-1.1

\textbf{Public Domain}: CC0-1.0, libtiff, Unlicense